# Confining photonic nanojet in a microwell on microsphere lens for highly efficient light focusing, signal amplification and quantitative detection

*Pengcheng Zhang[1]†, Bing Yan[2]†, Guoqiang Gu[1], Zitong Yu[1], Xi Chen[1], Zengbo Wang[2,*] and Hui Yang[1]\**

[1]*Laboratory of Biomedical Microsystems and Nano Devices, Bionic Sensing and Intelligence Center, Institute of Biomedical and Health Engineering, Shenzhen Institutes of Advanced Technology, Chinese Academy of Science, Shenzhen, China.*
[2]*School of Computer Science and Electronic Engineering, Bangor University, Dean Street, Bangor, Gwynedd LL57 1UT, UK.*
\*Contacting author: hui.yang@siat.ac.cn; z.wang@bangor.ac.uk

**ABSTRACT:** Dielectric microspheres or microbeads can squeeze light into the subwavelength scale via photonic nanojet (PNJ) focusing. This enables strong light-matter interactions within its focus and induces extraordinary effects such as enhancements in light emission, signal collection and various other applications. However, critical challenges exist on how to efficiently and precisely loading the studied objects into the desired location of the PNJ focusing, and subsequently measure the field and signal of interests precisely and reliably. Such technique is currently missing. We present for the first time a unique microwell-decorated microsphere lens ($\mu$-well lens), with a semi-open microwell sample compartment directly fabricated on top of the microsphere lens. The $\mu$-well lens confines PNJ in a semi-open microwell and allows passive trapping of individual micro-object into the PNJ focusing area with high efficiency and spatial accuracy. We demonstrate that individual fluorescent microsphere of different sizes can be readily introduced to the designated location with loading efficiency >70% and generates reproducible enhanced fluorescence signals with standard deviation better than that can be quantitatively measured. A comprehensive analysis on the optical properties of the $\mu$-well lens reveals the synergistic effect of field enhancement and collection efficiency on the optical enhancement. We finally employ this special microsphere lens for fluorescent-bead-based biotin concentration analysis. The results suggest a greatly enhanced sensitivity and highly improved detection limit, opening the door for its application in highly sensitive and quantitative detection. The $\mu$-well microsphere lens demonstrated here is promising for the development of next-generation on-chip signal amplification and quantitative detection systems with extremely high sensitivity and low detection limit;

it also provides a platform for investigating a wide range of light–matter interaction processes.

**INTRODUCTION**

The ability of confining and concentrating light energy into a very small volume permits the efficient delivering of high-intensity optical energy to the micro/nano objects of interests with high spatial resolution. This leads to enhanced light–matter interactions, which has a profound effect on the efficacy of various optical processes[1, 2]. It is of great importance for a variety of novel techniques in optical sensing, imaging and manipulating, with extensive applications in various fields, including precision bioassays[3, 4], cell biophysics[5], surface photocatalysis[6, 7], and surface enhanced Raman spectroscopy[8]. Undoubtedly, the ability to confine and concentrate light has emerged as a prominent interdisciplinary need and has attracted significant research interests. Conventional techniques for confining light into a deep subwavelength scale are widely performed by metal-based plasmonic structures including array of nanostructures or nanoparticles, by taking advantage of electromagnetic resonances on metals such as surface plasmon modes[9]. While these methods have certain advantages depending on the given application, there are still many common drawbacks such as intrinsic optical losses in metals, small mode volume and technically challenging nanofabrication processes, which hinders their practical applications.

Recent research trends now offer new approaches for squeezing light at the subwavelength scale via using dielectric microbeads including microsphere, microcylinder and others[10, 11, X1]. Compared with metallic structures, dielectric microbeads possess simple structure and almost no intrinsic absorption. Therefore, utilization of dielectric microbeads is considered as a simple and cost-effective route to focus and concentrate the light into subwavelength scale without requiring expensive nanofabrication facilities or complex near-field configurations[12, 13]. Upon illumination with the light, the dielectric microbeads can modulate the spatial structure of the incident light and converge into a concentrated light flux, which is termed as a "photonic nanojet" (PNJ). Such tiny light flux can propagate over several wavelengths along a path extending beyond the evanescent field region without any significant divergence, and maintain a subwavelength dimensions along the three-dimensional region. As the light is confined, its local field intensity can reach a value that is orders of magnitude higher than the incident intensity. Consequently, objects in the focus zone of the PNJ experience an enhanced electromagnetic excitation which stimulates strong light–matter interactions. This has resulted in the development of dielectric microbeads based optical techniques in various applications including super resolution imaging[14, 15], ultrasensitive optical detection[16-19], optical trapping and manipulation[20, 21]. However, delivering of PNJ to the targeted objects through dielectric microbeads are limited in two ways. First, the PNJ is in general generated in a free space perpendicular to the substrate containing dielectric microbeads with relatively small effective area. Thus, introducing the objects of interest into this area is rather difficult. Second, due to the heterogeneous distribution of the PNJ intensity field, signals that report the strength of light–matter interactions between the objects and PNJ could exhibit server fluctuations when spatial location of the object varies. This poses great challenges on achieving reliable signal amplification via PNJ and obtain accurate quantification and measurement as desired. We need a solution to position objects in desired location of PNJ, effectively, precisely, and reliably. This has not been achieved in the literature.

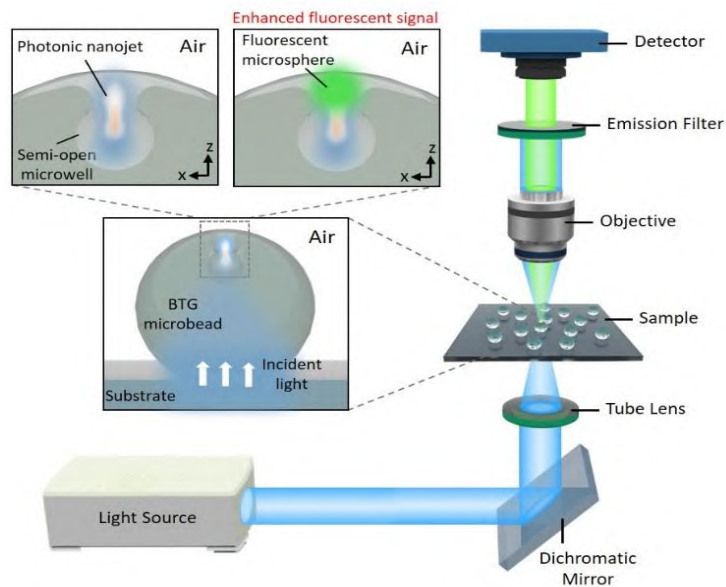

*Scheme 1. Schematic illustration of $\mu$-well microsphere lens, fabricated on barium titanate glass (BTG) microsphere with a laser drilled microwell, and the experimental setup designed for recording the signal of the fluorescent microspheres in the semi-open microwell. Individual fluorescent microsphere is loaded in the microwell. Upon illumination from the bottom of the BTG microbead, photonic nanojet is generated inside the microwell, which excites the fluorescent microsphere and induces enhanced light–matter interactions, leading to the emission signal amplification. The emission signal is collected from the top via transmissive optical configuration.*

In this study, we present for the first time an $\mu$-well microsphere lens which confines high-intensity PNJ in a semi-open microwell and allows passive trapping of single microscale object into the PNJ area with high degree of spatial accuracy (Scheme 1). The novel design greatly improves the efficiency of introducing the targets into the PNJ as well as suppresses their signal fluctuations, allowing reliably measurement of signals amplified by enhanced light–matter interactions. This paves the way to establish a new and highly sensitive quantitative detection approach. As a demonstration, we show fluorescent microspheres of different sizes can be efficiently introduced into desired PNJ area and their location in the PNJ can be readily controlled and adjusted. Consequently, their fluorescent signals are amplified by the unique design due to the enhanced light–matter interactions within the PNJ. Therefore, the fluorescent enhancement factor, which reports the strength of light–matter interactions between the fluorescent microspheres and PNJ, can be reliably and accurately measured. In our experiments, different optical configurations (reflection or transmission modes) were investigated and compared to understand its influences on the efficiency of signal amplification and acquisition process, with the goal to obtain optimized larger scale signal enhancement. Lastly, we employ this μ-well microsphere lens for fluorescence based biotin concentration analysis, which results in a 1.6-fold increase in sensitivity and 16-fold improvement on limit of detection, demonstrating the ability of quantitative and sensitive detection of targets.

**RESULTS AND DISCUSSION**

**1. Fabrication and characterization of μ-well microsphere lens**

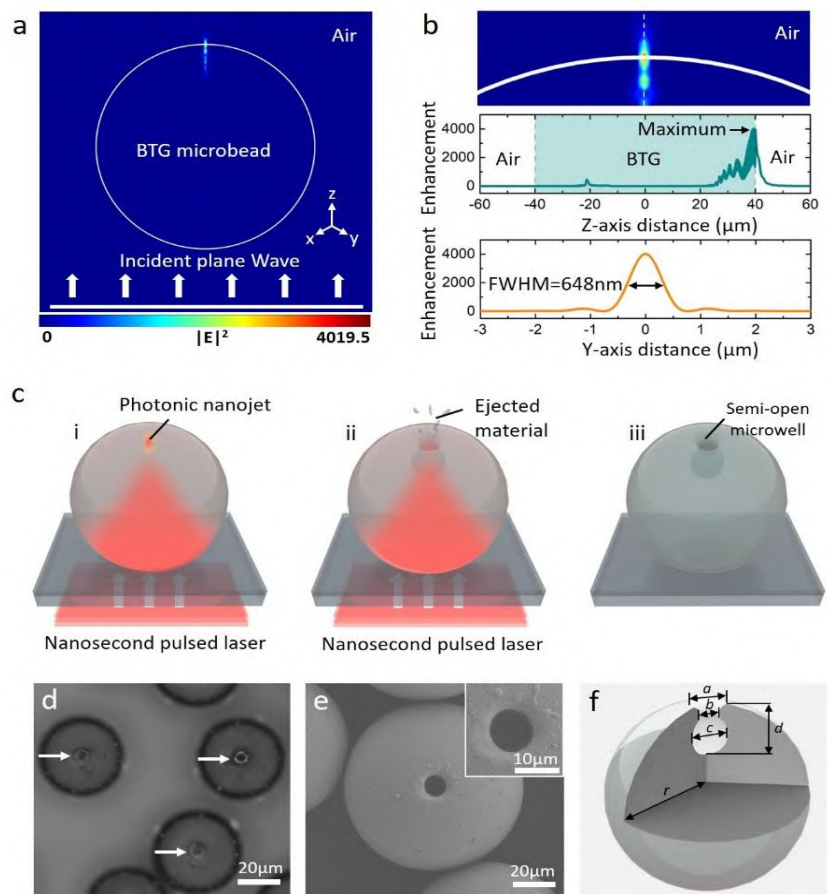

Figure1.(a) Calculated electric intensity field ($|E|^2$) focusing map by 80-um BTG microsphere upon illumination with plane wave of 1064nm wavelength. (b) Plot of the electric field intensity distribution along propagation z-axis and the full width at half maximum (FWHM) of the PNJ. The white and green areas in the middle image represent the area of air and BTG material, respectively. The maxima is shown to occur at the boundary of the BTG material and the air. (c) Schematic illustration of the fabrication process of the semi-open microwell in the BTG microbeads. (d) Photographic image of the as-prepared BTG microsphere with semi-open microwells. The white arrows indicate the semi-open microwells on each BTG microbead. (e) Scanning electron microscope (SEM) image of the as-prepared Microsphere lens. (f) Schematic illustration of the dimensions of the fabricated BTG microbeads with semi-open microwell.

Conventional barium titanate glass (BTG) dielectric microspheres (refractive index 1.9) with diameter of ~80 μm were utilized. The BTG microbeads were pre-deposited on an uncured polydimethylsiloxane (PDMS) spin-coated glass substrate to form a monolayer and were subsequently fixed by curing the PDMS layer. The particle density of the monolayered BTG microbeads can be tuned by controlling the number of deposition steps during the deposition process. Nanosecond pulsed laser was used to fabricate the semi-open microwell on the top of each BTG microbead. Based on the Mie calculation (Figure 1a), upon illumination with a plane wave (wavelength λ = 1064nm) from the bottom, a focused high-intensity beam, i.e., photonic nanojet (PNJ), is generated on the top surface of BTG microbead. The plot of the electric field intensity distribution along z-axis in Figure 1b shows that, an elongated PNJ with maximum enhancement factor about 4000 occurs on the boundary between the BTG microbead and the air. The amplitude of electric field intensity inside the BTG microbead raises sharply from around 10μm left off the boundary, then reaches a maximum at the edge of the boundary and rapidly decays outside the BTG microbead. Meanwhile, the PNJ exhibits a narrow waist with a full width at half maximum (FWHM) of 648nm (0.61λ). The high-intensity PNJ can generate heat that enables ablation and direct removal of BTG material in the focus area due to the thermal accumulation effect, leading to the formation of a semi-open microwell[22]. Besides, this approach can induce minimum damage around the microwell area. The fabrication process is illustrated in Figure 1c. Upon illumination from the bottom of the BTG microbead, the nanosecond laser beam (IPG nanosecond laser, central wavelength λ = 1064nm, pulse duration 100ns, pulse power up to 10w) as the heating pulse undergoes a focusing inside the BTG microbead due to the high refractive index of BTG material, generating a highly localized PNJ at the boundary on the rear side of the microbead (Figure 1c-i). The PNJ creates intensively localized energy at the boundary and causes the material to be ablated and removed, leaving a single semi-open microwell on the top of each BTG microbead (Figure 1c-ii and iii). By scanning the substrate with the nanosecond laser beam, large-scale array of BTG microbeads with semi-open microwell can be fabricated. It should be noted that, our microwell fabrication approach ensures that the generated semi-open microwell are accurately located at the top center of each BTG microbead, whose position is controlled by the incident beam angle. Detailed fabrication process can be found in supporting information.

These as-prepared microsphere lenses were characterized under the light microscope, as shown in Figure 1d. These semi-open microwells can be clearly observed since they exhibit dark circular area on the center of the top, arising from the scattering effect of the walls surrounding the semi-open microwell. The scanning electron microscopy (SEM) images under high magnification clearly show that a larger outer contour (~11μm) is on the top of the BTG microbead, Figure 1e. The diameter and height of the contour gradually decreases and eventually narrows into a semi-open microwell with a diameter of about 6μm. It is observed that the interior of the semi-open microwell is an ellipsoidal cavity with a maximum cross-sectional profile about 13μm and a height about 20μm, as illustrated in Figure 1f. Further details on the cross-section and the interior of the semi-open microwells are presented in supporting information (Figure SIXX). The size of the semi-open microwell can be controlled via tuning the power of the nanosecond laser beam and the number of laser exposure times during the fabrication process. Utilization of ultrafast laser, such as picosecond or femtosecond laser, would benefit the fabrication of semi-open microwells down to sub-micron scale.

## 2. Passive trapping of fluorescent microspheres into the PNJ inside microwell

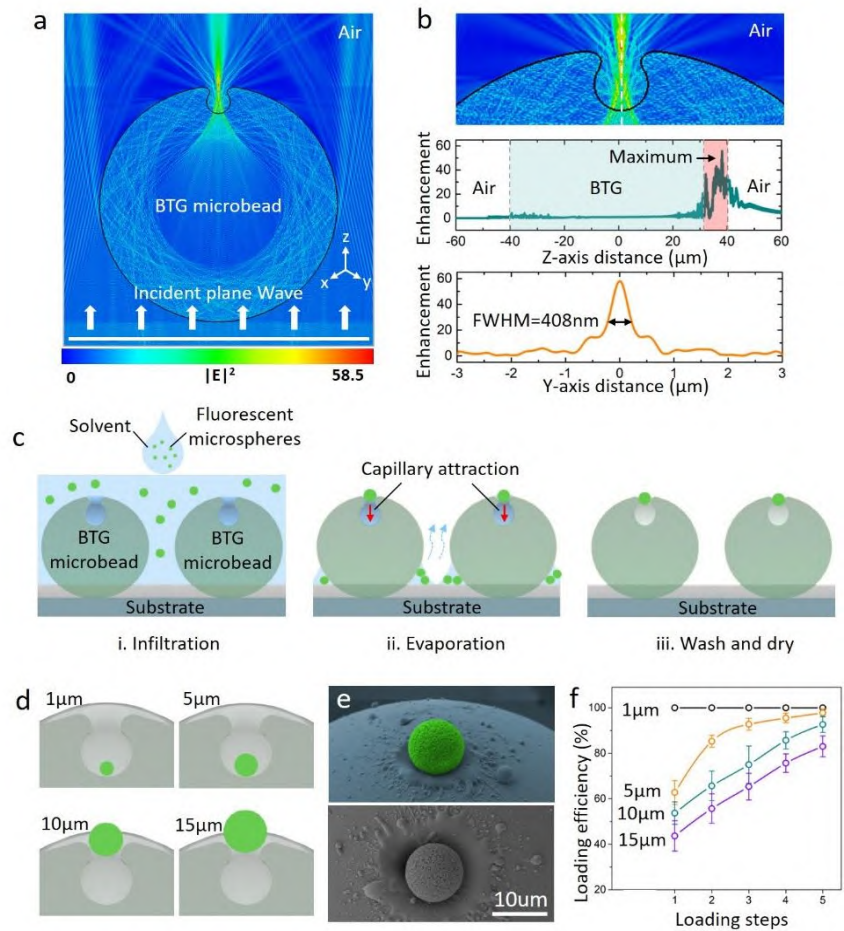

*Figure2. (a) Simulation based on finite- difference time-domain (FDTD) showing localized photonic nanojet can be generated inside the semi-open microwell. (b) Plot of the electric field intensity distribution along z-axis and FWHM of the PNJ. The white, green and pink areas in the middle image represent the area of air, BTG material and semi-open microwell, respectively. (c) Schematics depicting the process of the fluorescent microsphere trapping capability of the μ-well lens. (d) Schematic illustration of the positions of fluorescent microspheres with different sizes loaded in the semi-open microwell. (e) SEM images showing the fluorescent microsphere of 10μm located in the semi-open microwell. Upper: side view of the false color SEM image; Lower: top view of the SEM image. (f) Dependency of the loading efficiency on the number of loading steps. The full lines represent guides to the eye.*

Simulation based on finite- difference time-domain (FDTD) was performed to investigate the optical focusing property of the as-prepared μ-well lens in air ambient. The simulation shows that, upon illumination from the bottom, the incident light (plane wave with wavelength λ= 475nm, which is the excitation central wavelength of the fluorescent samples used later) can be converged and a photonic nanojet is generated inside the semi-open microwell, Figure 2a. Besides, the maximum of the electric field occurs inside the semi-open microwell and the lateral resolution of the photonic nanojet (FWHM) is 408nm, as shown in Figure 2b. The simulation indicates that, apart from the spherical or cylindrical optical structure, this optical semi-open microwell structure can also generate PNJ. It should be noted that, PNJs are usually developed in open sub-wavelength space for the commonly used spherical and cylindrical optical structure, which makes it technically difficult to accurately introduce target into this space. Current solutions such as random capture based on microfluidics or manually capture based on optical forces are severely hindered by their performance on throughput and the requirements on complexed experimental configurations[18, 21]. As a comparison, our $\mu$-well lens design in this study generates PNJ in a semi-open microwell, which provides a decent localized area to accommodate the targets. This implies that, once the target is loaded into the semi-open microwell, it can be simultaneously illuminated by the high-intensity PNJ without any further manipulations.

To demonstrate the microscale objects trapping capabilities, as well as measuring the signal boosted by enhanced light–matter interactions, fluorescent microspheres in aqueous dispersion were utilized. Previous studies have shown that substrates patterned by concave features possess local energy minima which can efficiently trap particles by employing self-complementary key-lock mechanisms in the solution during the assembly process[23]. Similarly, here we demonstrate that owing to the localized area provided by this $\mu$-well lens, the particles as target in dispersion can be passively trapped inside the semi-open microwell driven by the capillary force as the solvent gradually evaporates. Furthermore, as the carrier of the assembled fluorophores, fluorescent microspheres in the PNJ experience enhanced light–matter interactions and their emission intensity can be used to report the strength of the light–matter interactions. The target trapping process is illustrated in Figure 2c. Fluorescent microspheres with diameter of 1μm, 5μm, 10μm and 15μm were utilized respectively. A droplet of dispersion containing fluorescent microspheres (~1.5 × 10$^6$ ml$^{-1}$) is introduced to the surface of the fabricated lens using a micro-pipette. The dispersion covers the entire BTG lens array and infiltrates into the semi-open microwells. As the solvent of the dispersion gradually evaporates, the geometry of the semi-open microwell benefits a slower evaporation of the solvent, resulting capillary forces drawing fluorescent microspheres close by to the semi-open microwells[24]. Recessed circular geometry of the semi-open microwell benefits the onset of self-complementary key-lock mechanisms between microspheres and the semi-open microwells. Together with van der Waals, or depletion forces, fluorescent microspheres are confined and irreversibly trapped in the semi-open microwells. After gently washed with Milli-Q water, fluorescent microspheres left around the corner of the BTG microbeads can be washed away and eventually only microspheres in the semi-open microwells are remained.

Once the PNJ is generated, the trapped microspheres are automatically exposed in the enhanced electromagnetic field without further manipulation. Besides, due to the special geometry structure of the semi-open microwell, these fluorescent microspheres can exhibit different height distributions along the longitudinal direction, as shown in Figure 2d. Fluorescent microspheres with diameter (1μm and 5μm) smaller than the "neck" of the semi-open microwell (value b in Figure 1f) can enter the semi-open microwell while others (10μm and 15μm) are blocked and located outside the semi-open microwell. This suggests that, apart from introducing of the microspheres in the PNJ, the position of the microspheres along the longitudinal direction can be controlled by adjusting their sizes. SEM was used to characterize the fluorescent microspheres in the microwells after the sample loading procedure, Figure 2e. The SEM images confirm that, after the evaporation of solvent, the individual fluorescent microsphere with diameter of 10μm is targeted by the semi-open microwell. We performed multiple loading steps (each loading process in Figure 2c is seen as one step) and found that the loading efficiency depends on the number of loading steps and the size of fluorescent microspheres, Figure 3f. The loading efficiency increases as the number of loading steps increases. Typically, these semi-open microwells can be filled at a yield approaching 70% after 5 loading steps for all the fluorescent microspheres used in our experiment. However, given the same number of loading steps and the concentration of dispersion, higher loading efficiency is observed for smaller fluorescent microspheres. For example, nearly 100% loading efficiency is found for microspheres of 1μm while about 40% loading efficiency is found for that of 15μm after the first loading step. The size dependent loading efficiency is ascribed to the size induced loading preference during the assembly process[25, 26]. For the fluorescent microspheres with smaller size, they encounter higher possibilities to be trapped and meanwhile possess higher stability in the semi-open microwell, inducing higher loading efficiency compared with that of larger size. Unfilled semi-open microwells can be further eliminated by additional loading steps with a more dilute dispersion ($\sim 1.5 \times 10^5$ ml$^{-1}$). However, multiple fluorescent microspheres loaded in one single semi-open microwell or located beneath the BTG microbeads would bring interference on determination of the fluorescent signals in the following experiments. In order to eliminate this interference, it is necessary to load only individual microsphere in each single semi-open microwell, which can be achieved by employing a dilute dispersion and performing more loading steps. Here, samples for single microsphere loading in individual microwells were utilized in the following experiments. It should be noted that, unbalanced capillary force during evaporation or the electrostatic force may cause some fluorescent microspheres (<10% for fluorescent microspheres of 1μm and 5μm) to randomly adhere to the inner wall surface of the semi-open microwell. This leaves them at an off-center position, which can be distinguished by reading their focal length or measuring their geometric position to the center. Microspheres whose horizontal position deviated from the center by more than 50% (approximately equal to FWHM of the PNJ divided by 1μm) of their diameter were excluded from the statistics. Such phenomenon was rarely observed on microspheres located outside the semi-open microwell, largely due to the strong self-complementary key-lock mechanisms with the "neck" of the semi-open microwell. Detailed description of the loading process is presented in the supporting information.

The loading strategy here provides a simple yet effective approach to introduce the individual microsphere into the PNJ with controlled position. Besides, the self-complementary key-lock mechanisms enables the physical contact between microspheres and the semi-open microwells. Thus, the spatial variations of the microsphere in the semi-open microwell are highly reduced. As a result, microspheres of the same size are located at the same height along the longitudinal direction and thus experience nearly identical intensity of the PNJ, leading to suppressed signal fluctuation. This benefits the reliably measurement of signals boosted by enhanced light–matter interactions and enables its application in quantitative detection.

## 3. Reliable and accurate optical enhancement and signal quantification via measuring the fluorescent enhancement factor

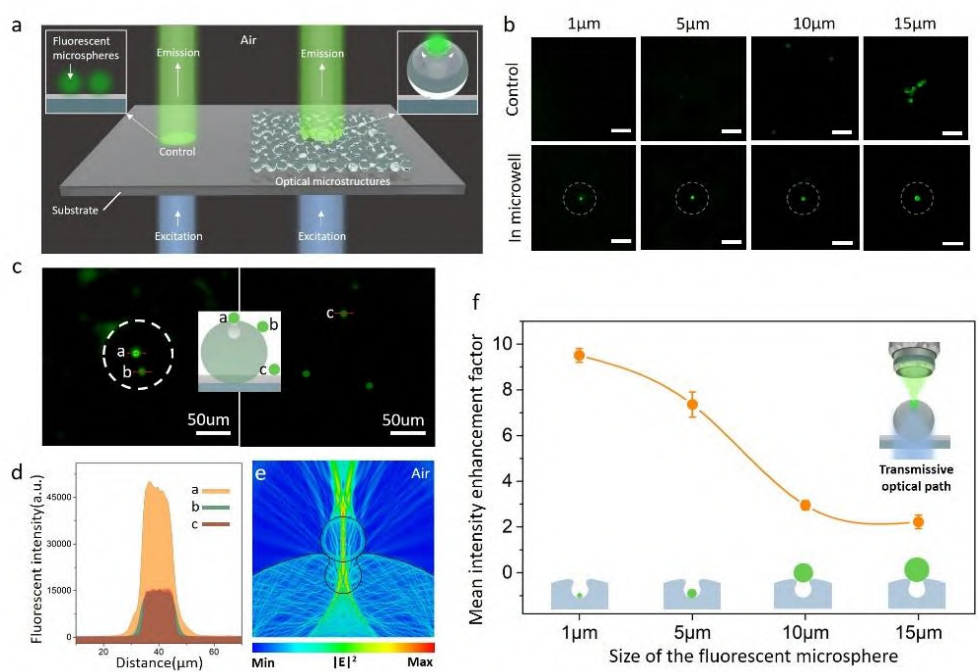

*Figure3. (a) Schematic illustration of the signal recording for the fluorescent microspheres in the semi-open microwells and on the substrate (control) and (b) corresponding recorded fluorescent signal. (c) Optical images of the fluorescent signal in the microwell, on the vicinity of the microwell and on the substrate and (d) the intensity of the corresponding fluorescent microspheres showing the difference of the fluorescent signal explicitly. (e) Simulation based on finite- difference time-domain (FDTD) showing the impact of a microsphere on the PNJ. (f) The enhancement factor v.s. the size of the fluorescent beads measured by experiment (error bar stands for standard deviation).*

After loading, the samples were gently washed with Milli-Q water and then were completely dried with nitrogen. Their fluorescent signal was collected in the air ambient under the optical configuration as shown in Scheme 1. The excitation light (central wavelength λ=475nm) is illuminated from the bottom of the BTG microbeads, inducing PNJ generated in the semi-open microwell on the top. The emission light (central wavelength λ=525nm) from the fluorescent microspheres in the semi-open microwell was collected by an objective (10×, numerical aperture, NA, =0.25) above the samples. To prove the optical enhancement of this $\mu$-well lens, the emission light from the fluorescent microspheres on the substrate was also collected in the identical excitation power and exposure time as a control (Figure 3a). As expected, fluorescent microspheres inside the semi-open microwells exhibited higher emission intensity than that on the substrate, as shown in Figure 3b. Bright visible emission was observed for the fluorescent microspheres inside the semi-open microwells, while only inconspicuous emission was captured for that on the substrate. The contrast on their emission intensity can be clearly observed in Figure 3c. The data of emission intensity extracted from Figure 3c shows that more than 3-fold enhancement on emission intensity was obtained for the fluorescent microspheres in the semi-open microwell, compared with that located at the vicinity of the semi-open microwell or on the substrate (Figure 3d). These results indicate that the generated PNJ in the semi-open microwell possesses higher electric field intensity that significantly exceeds the illuminating excitation light. This has also been confirmed in the simulation in Figure 3e, where the external electromagnetic field experienced by fluorescent microsphere is tailored by the $\mu$-well microsphere lens by generating PNJ, which delivers high-intensity optical power to the fluorescent microsphere.

To quantify the optical enhancement capabilities of this $\mu$-well lens, we define mean emission intensity enhancement factor $\eta_M$ as

$$\eta_M = \frac{\overline{I_\mu}}{\overline{I_c}}$$

(1)

, where $\overline{I_\mu}$ represents the mean emission intensity of the fluorescent microspheres when positioned inside the microwell, and $\overline{I_c}$ represents the mean emission intensity of the fluorescent microspheres on the glass substrate (control experiment). The measured magnitude of $\eta_M$ reflects the optical enhancement capability of this $\mu$-well lens, while the measured standard deviation (SD) represents the signal fluctuation, i.e., the stability of the enhanced emission signal (shown as error bar in plot). We repeated measurements of $\eta_M$ for different sized fluorescent microspheres at different locations, and statistical results were shown in Figure 3f. Note the detected emission signal from individual fluorescent microsphere presented here was randomly selected, and consistent uniformity can be found for all the fluorescent microspheres with the same size. Overall, the fluorescent signal turned to be quite stable for all measured cases, with SD all below 7.5%. This in turn suggests the developed $\mu$-well systems are overall stable and reliable for precision measurement of micro/nanoscale signals. A slightly larger SD values were observed for smaller particles of sizes 1 and 5μm, when compared to 10μm and 15μm sized particle. This is caused by the fact that smaller fluorescent particles are more likely to be deposited at different locations inside the microwell. On the other hand, Figure 3f also shows that the enhancement factor $\eta_M$ decreases with particles sizes. For instance, up to one order (~9.5-fold) of enhancement was obtained for the 1-μm-diameter fluorescent microspheres, about 2-fold of enhancement for the 15μm fluorescent particle. Since differently sized fluorescent microspheres could be trapped at different z-locations in the microwell, it would lead to a general 'size effect' on the enhancement factor for the fluorescent signal, i.e., the smaller the particle size, the higher the enhancement factor. The focusing jet inside microwell (FWHM: 648 nm, see Fig.1b) is in comparable size to 1 µm particles, but smaller than other size of particles of 5, 10 and 15 µm. This suggests for larger particles the nanojet doesn't provide a full illumination of the particle, but only partially. The excitation of fluorescence molecules will be limited to the excited regions determined by nanojets (i.e. similar level of input energy) while the emitted fluorescent light will distribute through the whole sphere (i.e., larger area output). This will cause a reduced average intensity and thus a lower enhancement factor as observed in experiments.

## 4. Influence of optical configurations on the efficiency of signal acquisition

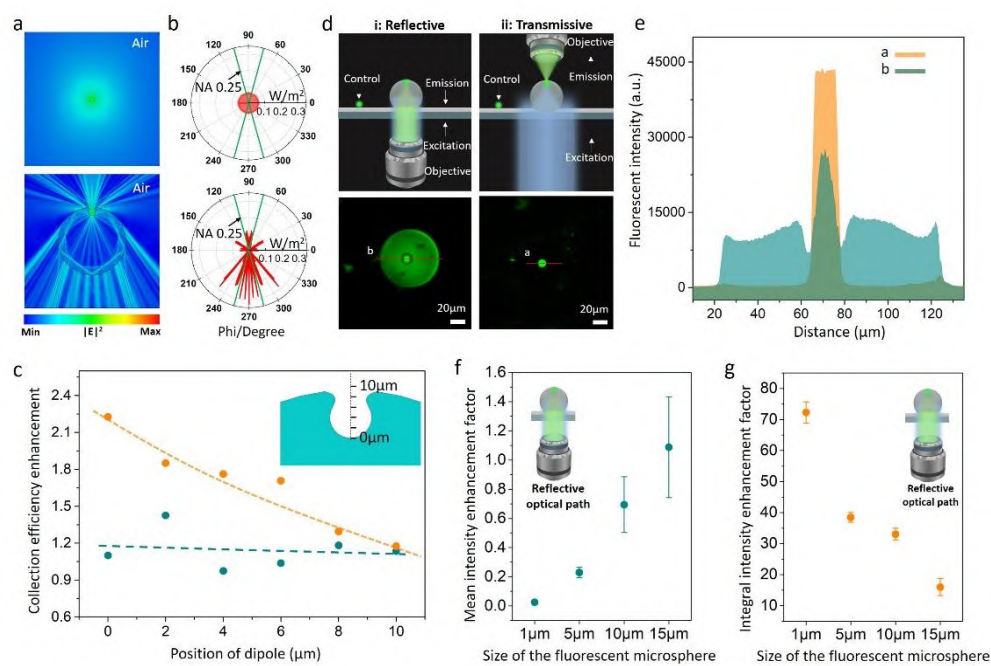

*Figure4. (a) Simulations based on FDTD showing the far-field emission collection efficiency for a dipole source in the presence or absence of the µ-well lens. The dipole is in a free space for the upper image and is set on the bottom of the semi-open microwell for the lower image. (b) The corresponding radiation pattern showing the effect on radiation pattern by this µ-well lens. The crossed green lines indicate the light collection area under the objective with numerical aperture (NA) 0.25 used in our experiments. (c) Calculated far field collection efficiency enhancement on the utilization of µ-well lens. The dipole is set at the different z-positions above the bottom of the semi-open microwell, as the value indicated on the abscissa. The green dots represent the enhancement while collected from the position of 90 degree and the orange dots from the position of 270 degree, as shown in b. The dashed lines represent guides to the eye. (d) Schematic illustration of two different optical configurations used for comparing their influence on fluorescent signal collection (upper images) and the corresponding obtained fluorescent images (lower images). i: transmissive optical path for collecting the emission signal directly from the top (90 degree); ii: reflective light path for collecting the emission signal after passing through the µ-well lens from the bottom (270 degree). For both configurations, fluorescent microspheres on the*

*substrate are used as the control. (e) The fluorescent intensity of the corresponding lower fluorescent images in (d). Experimental results of the mean emission intensity enhancement factor (f) and integral intensity enhancement factor (g) for fluorescent microspheres with different sizes collected under the optical configurations in (d)-ii.*

Besides considering the excitation of fluorescent particles, we also considered how the fluorescent signal was collected by the u-well lens and propagate to the far-field detector using dipole scattering FDTD simulation (Fig.4), i.e., the influence of far-field light collection efficiency on the size dependent enhancement factor. It has been reported that dielectric microbeads could enhance the efficiency of far-field light collection by converting the wavefront of a point light source to quasi-plane-wave[16, 19]. Indeed, our $\mu$-well lens does this as shown by comparative FDTD results in Figs. 4a and 4b. In Fig. 4a, a dipole source is radiating energy uniformly into its surrounding space, while in Fig. 4b one can see the u-well lens significantly change the radiation directivity of the dipole source and redirects the energy mostly toward the top (90-deg angle) and bottom directions (270-deg angle) of the lens, leading to significant increase in signal collection efficiency.

In our experiments, the emission source is excited fluorescent microsphere whose position could be varied at different z-axis in the microwell. In the simulation, we set the dipole source location at different z-positions above the bottom of the semi-open microwell and the corresponding efficiency of far-field light collection is calculated (Figure 4c). The results show that the light collection efficiency does not change significantly with the position of dipole, with a mean collection efficiency enhancement value of around 14%. This implies that utilization of this $\mu$-well lens only leads to minor influences on $\eta_M$ while collecting the fluorescent signal from the top direction as in our experiment. Based on these discussions, a simple model that includes these influencing factors is developed to explain how the mean emission intensity enhancement factor evolves with the size of the targets, which is presented in supporting information.

Note the simulation results in Figure 4c shows significant increase of as high as 2.3-fold in dipole far-field collection efficiency while collected from the bottom of the $\mu$-well lens (direction of 270 degree in Figure 4b). It suggests the possibility to further increase the emission enhancement through this $\mu$-well lens via utilizing reflective optical configuration (RL configuration) as illustrated in Figure 4d-i. In reflective optical configuration, the excitation light is illuminated from the bottom of the $\mu$-well lens, inducing the localized PNJ that excites the fluorescent microsphere, the same as that in transmissive optical configuration (TR configuration, Figure 4d-ii.). However, quite different from the emission collection in TR configuration, where the emission is directly collected via objective above, in the RL configuration the emission is collected after passing through the $\mu$-well lens beneath. According to the simulation results (Figure 4c), significantly increased energy is guided toward the bottom direction after modified by the $\mu$-well lens, it is expected more emission is collected. To validate this possibility, we recorded the emission signal using RL configuration in Figure 4d-i. Under the RL configuration, emission from the fluorescent microsphere is modified and projected by the $\mu$-well lens. The emission is then imaged and collected via focusing on the contour of the dielectric microbeads, which shows bright emission in a large circle area under view, as shown in the lower image of Figure 4d-i. The bright spot in the center arises from the scattering effect of the interior walls of the semi-open microwell. As a comparison, emission form the same fluorescent microsphere is also recorded under TR configuration and their profile of fluorescent intensity is shown in Figure 4e. It is found that, compared with TR configuration, the fluorescent intensity exhibits low fluorescent intensity but higher value on integral intensity for RL configuration. To investigate the emission enhancement effect of the $\mu$-well lens under RL configuration, we quantified the mean emission intensity enhancement factor $\eta_M$ by comparing the mean emission intensity of the fluorescent microspheres in the semi-open microwell and on the substrate, same as equation (1). The calculated $\eta_M$ under RL configuration is shown in Figure 4f. Different form the trends in TR configuration, $\eta_M$ in RL configuration increases with the increasing size of the fluorescent microspheres. However, the value of the enhancement factor is much lower compared with that under TR configuration. The value of $\eta_M$. is less than 1 for the fluorescent microspheres of 1µm,5µm and 10µm, and slightly larger than 1 (1.09±0.34) for the that of 15µm. For instance, the $\eta_M$ is found to be only 0.03±0.01 for fluorescent microspheres of 1µm in RL configuration, while $\eta_M$ is 9.51±0.30 for that in TR configuration. The results show that utilization of RL configuration does not effectively increase the detected mean emission intensity. On the contrary, the detected mean emission intensity from the fluorescent microspheres in the semi-open microwell is even lower than that on the absence of this $\mu$-well lens for smaller fluorescent microspheres. Given that the recorded emission signal is from single fluorescent microsphere in each semi-open microwell, as discussed in the target trapping process, we assign the observed phenomenon as "dilution effect" resulting from the projection of emission signal on larger area in RL configuration. As shown in the lower image of Figure 4d-i, emission signal from individual fluorescent microsphere is distributed and imaged on a larger area (i.e., the cross-sectional area of the dielectric microbead) than the actual size of the fluorescent microsphere, causing the dilution of emission intensity. The degree of dilution depends on the size ratio between the fluorescent microsphere and the dielectric microbead, with lower degree of dilution occurring on the large microspheres. Thus proportional relationship was observed between the fluorescent enhancement factor and the size of fluorescent microsphere. Therefore, under RL configuration, $\eta_M$ can be increased by adopting fluorescent microsphere of larger size. However, since fluorescent microspheres of larger size are not favored for stably residing on the semi-open microwell, it is not feasible to introduce larger fluorescent microspheres into the semi-open microwell. Another approach to increase $\eta_M$ is to increase the amount of fluorescent microspheres (i.e., fluorescent microspheres of 1µm and 5µm) loaded in the semi-open microwell. Since the detected and imaged area under RL configuration is constant (equal to the cross-sectional area of the dielectric microbead), more fluorescent microspheres in the semi-open microwell can lead to the accumulated emission and higher mean intensity, resulting a higher $\eta_M$. Indeed, it is found that, loading of more fluorescent microsphere can lead to $\eta_M$ as high as 200 (FigureSI), which is consistent with the results from the previous report under RL configuration[19]. However, since it is technically difficult to precisely control the number of microspheres loaded into the semi-open microwell during target loading process, the $\eta_M$ amplified by this approach under RL configuration is shown to vary with a relative large deviation, hindering its application in quantitative detection.

Although the mean emission intensity is reduced, the detected integral emission intensity under RL configuration is pronounced increased. We define the integral intensity enhancement factor $\eta_I$ as

$$\eta_I = \frac{\int I_u d\Omega}{\int I_c d\Omega}$$

(2)

,where $\int I_u d\Omega$ represents the detected integral emission intensity of individual fluorescent microsphere in the microwell after passing through the dielectric microbeads. $\int I_c d\Omega$ represents the integral emission intensity of individual fluorescent microsphere on the glass substrate. The dependency of the integral intensity enhancement factor on the size of the fluorescent microspheres in show in Figure 4g. It is found that, up to 75-fold enhancement on the integral intensity can be achieved on the fluorescent microsphere of 1μm. The integral intensity enhancement factor $\eta_I$ is observed to have an inverse proportional relationship with the size of fluorescent microspheres, which arises from the synergistic effect of PNJ enhancement and enhanced light collection efficiency. Since the enhancement of PNJ and light collection efficiency are both inversely proportional to the size of the fluorescent microsphere (Figure 3f and Figure 4c, respectively), the integral intensity also exhibits inverse relationship with the size of fluorescent microspheres. Owing to its capability to effectively enrich the targets and collect the fluorescence, this RL configuration can be used to detect tiny objects of low fluorescence beyond the resolution of the microscope by analyzing the distribution of light field collected, as reported by previous study[16, 27, 28].

## 5. Application of the $\mu$-well lens on biotin concentration analysis

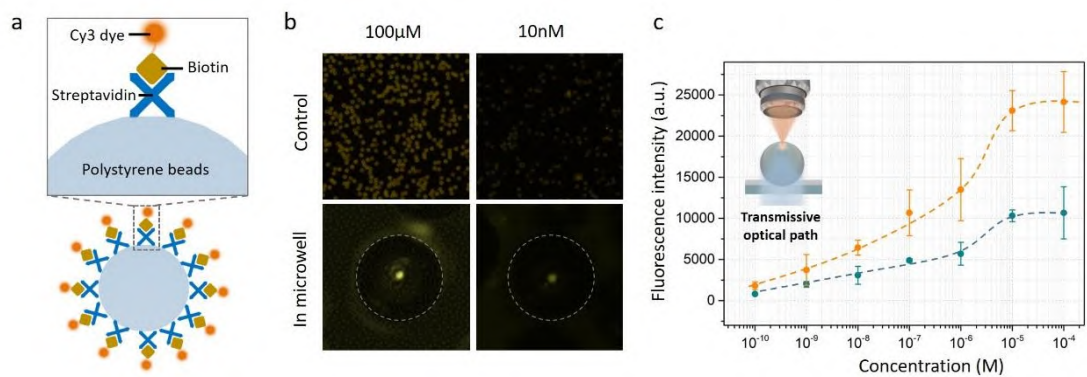

Figure5. (a) Schematic illustration of the cy3 labeled polystyrene (PS) microbeads. (b) Recorded

*fluorescent signals of the fluorescent dyes labeled PS microbeads in the semi-open microwells and on the substrate (control). (c) Recorded fluorescent intensity in different concentrations of biotin under transmissive optical configuration. The orange and blue dots represent the intensity recorded with and without fluorescent signal amplification, respectively. The dashed lines represent guides to the eye.*

As one of the possible practical applications of this $\mu$-well lens with the capability of signal quantification, we demonstrated a fluorescent bead-based quantitative biotin concentration analysis by employing this $\mu$-well lens under TR configuration. Polystyrene (PS) microbeads of 5μm were modified with Cyanine 3 (Cy3) dye labeled biotin with different concentrations, ranging from 100μM down to 100pM, Figure 5a. Fluorescent signal at the focal plane of the modified PS microbeads in the semi-open microwell and on the glass substrate was collected under TR configuration. The corresponding recorded fluorescent images in concentration of 100μM and 10nM are shown in Figure 5b, respectively. It is shown that utilization of this novel $\mu$-well lens pronounced increases the fluorescent intensity of the modified PS microbeads. Meanwhile, the fluorescent intensity under different concentration of biotin is shown in Figure 5c. It is observed that, under the same power of excitation in TR configuration, the fluorescence intensity of the PS microbeads located in the semi-open microwell is higher than that in the absence of the $\mu$-well lens at each concentration. The higher slope extracted from Figure 5c for the PS microbeads in the semi-open microwell implies an increase in the sensitivity of biotin response upon signal amplification (Figure SI). About 1.6-fold increase in sensitivity was obtained. Moreover, the amplification on the fluorescent signal improves the signal-to-noise ratio (SNR) and enables a lower limit of detection (LOD). For instance, at a concentration of 0.1 nM (corresponding to 0.024ng/ml), fluorescent signal amplified by the $\mu$-well lens is still detectable with SNR=9.0, while the fluorescent signal without amplification cannot be distinguished from the background noise. By assuming a linear relationship between the fluorescent intensity and the concentration of the biotin (in logarithmic scale), we deduced a LOD of 18pM (corresponding to 4.5pg/ml) with SNR=3.0 upon signal amplification (Figure SI). Therefore, the utilization of our novel $\mu$-well lens for signal amplification improved the LOD ~16 times relative to that without amplification (from 280pM to 18pM with SNR=3.0).

**CONCLUSION**

Method of inducing reliable optical enhancement and achieving accurate signal quantification through photonic nanojet is previously undescribed. In this work, we have taken an important step in validating the method by employing a dielectric microsphere with an microwell design. We have demonstrated that confined photonic nanojet can be generated in a semi-open microwell and individual fluorescent microsphere can be efficiently introduced into the PNJ area with high efficiency and spatial accuracy. This has led to reliable optical enhancement and quantitative signal measurements. In addition, a comprehensive analysis on the optical properties of this $\mu$-well lens reveals the mechanism of the optical enhancement, providing possibility of detecting objects of low fluorescence beyond the resolution. The ability of this $\mu$-well lens to offer simple yet efficient trapping capability for microscale objects and to provide reliable optical enhancement enabling quantitative signal measurements make it useful in the field of highly sensitive optical detection. Besides, this $\mu$-well lens also exhibits great potential to be used as a platform to study a wide range of light–matter interaction processes.

**Acknowledgment:**
BY and ZW acknowledge funding support from (1) European Regional Development Fund (ERDF), project 'Center for Photonics Expertise', Grant number 81400. (2) Royal Society, UK-China international exchange grant, No: IEC\NSFC\181378.